\documentclass[amssymb,amsmath,aps,twocolumn,nofootinbib]{revtex4-1}
\usepackage{natbib, graphicx, subfigure, subeqnarray,fancyhdr,amsmath,epstopdf,multirow,color, amssymb, cancel}
\usepackage{natbib}
\usepackage{graphicx,subfigure,subeqnarray,fancyhdr,amsmath,epstopdf,multirow,color, mathrsfs,amsmath,amssymb, natbib,setspace}
\usepackage[colorlinks=true, citecolor=red, linkcolor=blue,urlcolor=blue ]{hyperref}

\newcommand\uvec[1]{\hat{\mathbf{{#1}}}}

\newcommand{\pe}{_{\perp}}
\newcommand{\pa}{_{\parallel}}

\def\k{\hat{\bf k}}
\def\N{\hat{\bf N}}

\def\Ub{{\bf U}_b}
\newcommand{\Omegab}{{\boldsymbol{\Omega}}_b}

\newcommand{\omegaf}{{\boldsymbol{\omega}}}
\newcommand{\sN}{\sum\limits_{i=1}^{N_f}}

\newcommand{\putindeepbox}[2][0.7\baselineskip]{{%
    \setbox0=\hbox{#2}%
    \setbox0=\vbox{\noindent\hsize=\wd0\unhbox0}
    \@tempdima=\dp0
    \advance\@tempdima by \ht0
    \advance\@tempdima by -#1\relax
    \dp0=\@tempdima
    \ht0=#1\relax
    \box0
}}

\begin{document}
\title{Swimming of peritrichous bacteria is enabled by an elastohydrodynamic instability}
\author{Emily E. Riley
}
\altaffiliation{Present address: Centre for Ocean Life, Technical University of Denmark}
\thanks{These authors equally contributed to this work.}
\author{Debasish Das}
\thanks{These authors equally contributed to this work.}
\author{Eric Lauga}
\email{e.lauga@damtp.cam.ac.uk}
\affiliation{Department of Applied Mathematics and Theoretical Physics, University of Cambridge, UK.}
\date{\today}
\begin{abstract}

Peritrichously-flagellated bacteria, such as {\it Escherichia coli}, self-propel in fluids   by using  specialised  motors to rotate multiple helical filaments.  The rotation of each motor is transmitted to a short flexible {segment called  the hook} which in turn transmits it to a flagellar filament, enabling  swimming of the whole cell. Since multiple motors are   spatially distributed on the body of the organism,   one would expect the propulsive forces from the filaments to  push against each other leading to negligible swimming. We use a combination of computations and theory to show that   the swimming of multi-flagellated bacteria is enabled by an elastohydrodynamic bending instability occurring for hooks more flexible than a critical threshold. Using past measurements of hook bending stiffness, we demonstrate how the design of real bacteria allows them to be safely on  {the side of this instability that} promotes systematic swimming.

\end{abstract}
\maketitle


Although out of sight, bacteria dominate chemical processes on our   planet.  They  are   the most abundant  organisms on earth and, equipped  with the ability to live in extreme and hostile conditions, they play crucial   roles  in both the environment  and     human health~\cite{BiologyBook}.   
Many bacteria self-propel in response to physical and chemical cues by actuating  specialised, rotary  motors in bulk fluid environments~\cite{berg03}. Each  motor imposes a moment normal to the surface of the cell body  transmitted to a helical flagellar filament via a short elastic segment called    the hook  that acts  as a universal joint~\cite{Block1991,Samatey2004}.   
Due to  the helical nature of flagellar filaments, the rotation imposed by each motor is not time-reversible and as a result bacteria are able to swim~\cite{Purcell1977}.  While a flagellar filament can take one of  eleven polymorphic forms, the normal   form used for swimming is left-handed and rotates counter-clockwise (CCW, looking from the flagellum to the cell) propelling the bacterium cell-first~\cite{Turner2000,Darnton2007}, a  type of swimmer known as  a pusher~\cite{Powers2009}. If  the same left-handed helix were to rotate in the opposite direction then the cell would swim flagella first and be a puller~\cite{Murat2015}.

\begin{figure}[b]
\centering
\includegraphics[width=0.39\textwidth]{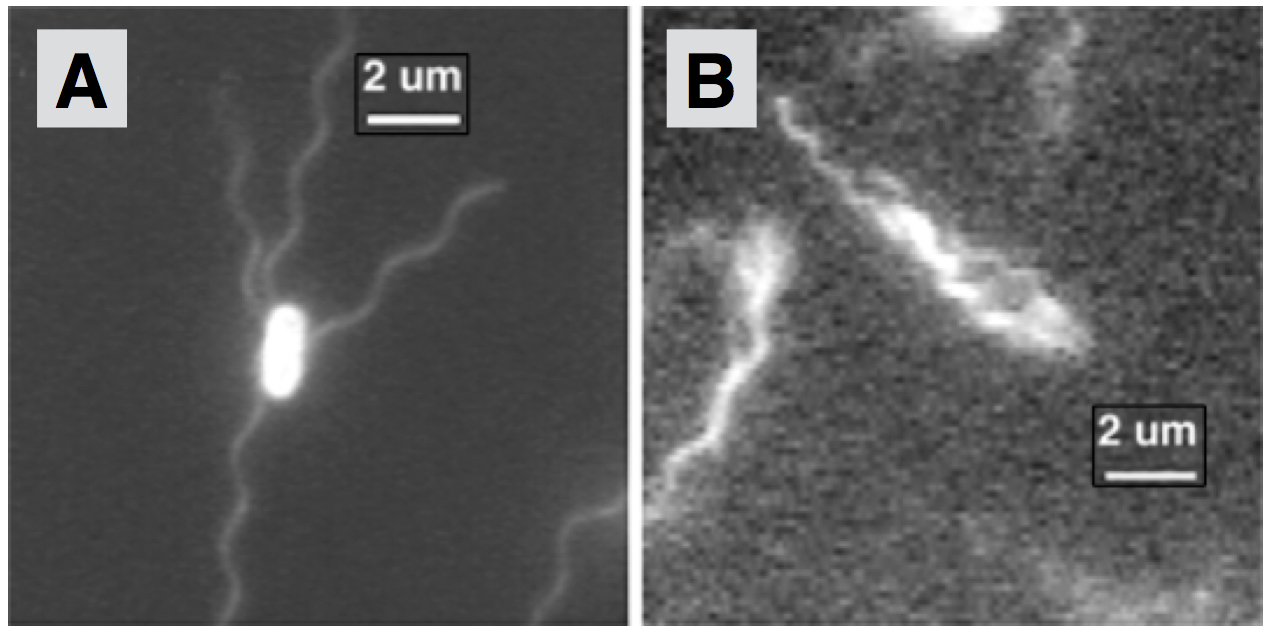}
\caption{Swimming \emph{E.~coli} bacteria. A: Peritrichous  bacterium with multiple flagella spatially distributed around the cell body; 
B: Flagellar filaments are located behind  the swimming cell in a helical bundle whose rotation push the cell forward. Reproduced from Turner, Ryu \& Berg (2000) {\it J. Bacteriol.}, {\bf 182}, 2793--2801 \cite{Turner2000}. Copyright 2000 American Society for Microbiology.}
\label{Fig1bergbact}
\end{figure}

Peritrichous  bacteria possess multiple flagella that can grow   from essentially any point on the cell body surface~\cite{Gutten2012, Ping2010}. Well-studied examples include \emph{Escherichia coli} (\emph{E.~coli}, Fig.~\ref{Fig1bergbact}A), \emph{Bacillus subtilis} and  {\emph{Salmonella enterica}}. During the swimming of these pusher cells, all flagellar filaments  gather and bundle at one end of the body propelling the cell forward  (Fig.~\ref{Fig1bergbact}B).   The main advantage of possessing multiple flagella is not increased propulsion~\cite{Mears2013} but rather  the ability to change direction  via tumbling. This occurs when at least one  of the rotary motors slows down~\cite{Scharf2002} or reverses its direction~\cite{Darnton2007} causing the bundle to break-up (unbundling). At the end of a tumble,  the motors return to their swimming state, the bundle reforms and the bacterium swims in a new direction. Through a modulation of the tumbling frequency, bacteria can move towards   favourable environments~\cite{BergBook}. 
Crucial for the formation of the filament bundle, and successful swimming, is the flexible hook. When the hook is stiffened, bacteria are   stuck in a tumble mode and can barely swim~\cite{Brown2012}.  
Hook flexibility is also crucial for singly flagellated bacteria, enabling   changes in swimming directions via buckling~\cite{Son2013} but causing unstable locomotion if it is too flexible~\cite{Shum2012}.

Much theoretical work has been devoted to predicting  the   propulsion mechanisms of self-propelled bacteria  \cite{LaugaReview}.  Most   studies assume a fixed relative position between  helical filaments (or  bundles) and the cell body, and modelling tools have been developed to address both    swimming \cite{Keller1976,Powers2009,Hyon2012,Rodenborn2013} and the bundling/unbundling process \cite{Reigh2013,Kanehl2014}.

If peritrichous  bacteria have similar    filaments  distributed  spatially around their cell body, why are the flagella not  all pushing against each other leading to negligible swimming? In this paper we use a combination of computations and theory to show that swimming is enabled by  an elastohydrodynamic instability of the hook. If  hooks are too rigid,     flagellar filaments always point normal to the cell body surface    and never bundle.  In contrast, when the bending rigidity of the hook    is below a critical threshold, the  feedback between the flow induced by the flagella and  hook bending leads to a   conformational instability resulting in all flagellar filaments  gathered at the back of the cell, and  net locomotion. 
This sharp transition from negligible   to successful swimming  is observed numerically  with decreasing hook stiffness and  we show that  this instability can  be  rationalised using a simple model of a cell propelled by two straight active filaments.   By examining past measurements of hook flexibility, we  demonstrate that bacteria are safely designed to be on the swimming side of the instability.

\section*{Results}
\subsection*{Modelling of multi-flagellated bacteria}
\begin{figure}[t]
\centering 
 \includegraphics[width = 0.41\textwidth]{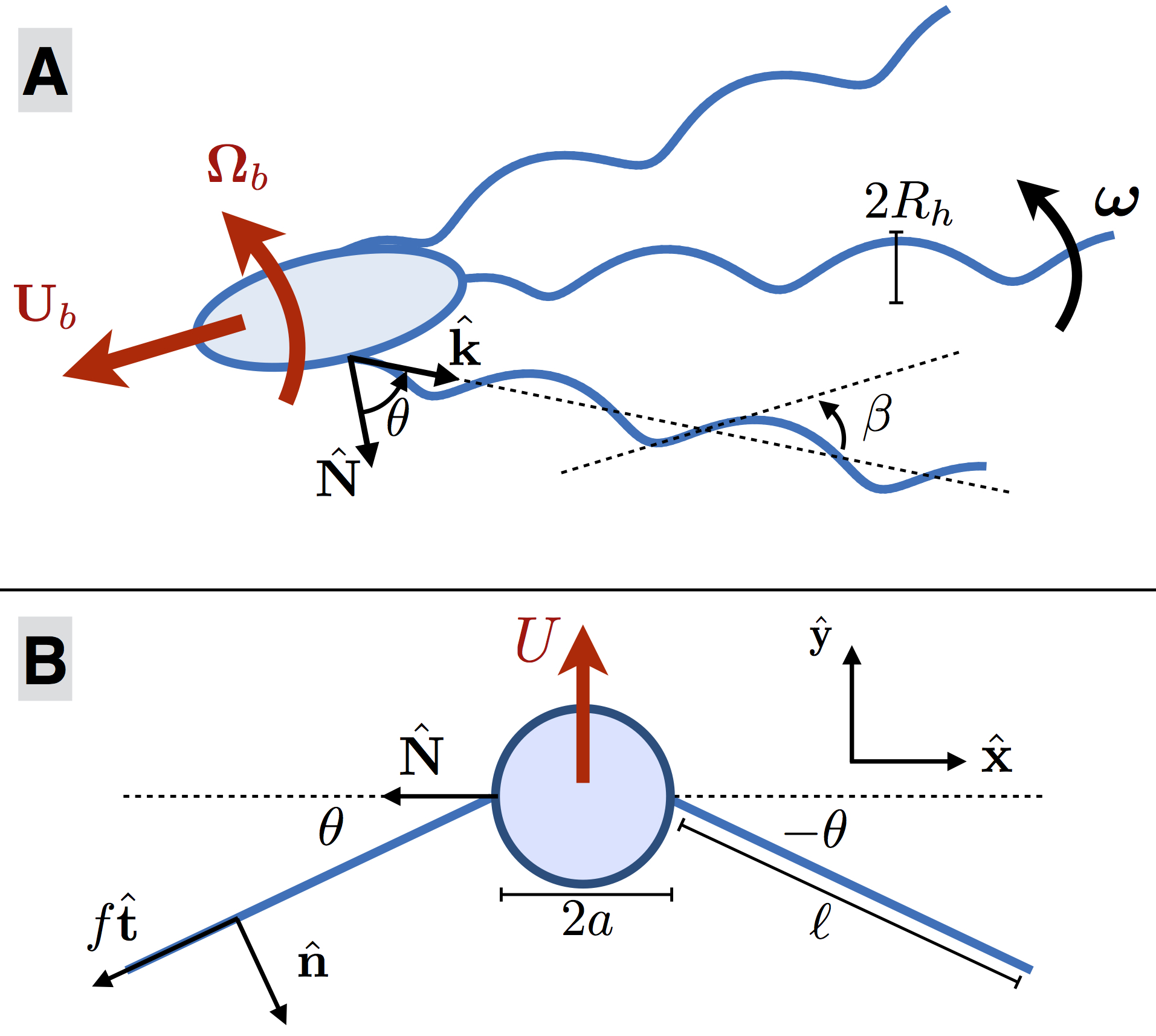}
\caption{A: Computational model of a peritrichous  bacterium   actuating $N_f$ helical filaments (radius $R_h$; angle $\beta$) by rotating them about their axis $\k$ with prescribed angular velocity. The flexible hook  acts elastically  to align the helix axis with the normal 
 $\N$ to the cell body. 
B: Simplified model to capture the elastohydrodynamic instability. Two straight active filaments  of length $\ell$   attached  on either side of a spherical body of radius $a$  are tilted at an angle $\pm\theta$ away from the cell body surface normal, $\N$,  and act on the cell  with tangential force $f\ell\uvec{t}$  resulting in  swimming of the model bacterium  with velocity $U\uvec{y}$.}
\label{BactDiagram}
\end{figure}

We start by building a computational model of the locomotion of a peritrichous  bacterium, as outlined in the Methods section with mathematical details in supplementary information \footnote{Supplementary information  available upon request  by writing to: \href{mailto:e.lauga@damtp.cam.ac.uk}{e.lauga@damtp.cam.ac.uk}.}.  We  
consider a   bacterium propelled by $N_f$ flagella  (Fig.~\ref{BactDiagram}A) with a cell body in the shape of  a prolate ellipsoid. Each flagellum consists of: (i) a rotary motor that generates a fixed rotation rate about the axis of the flagellar filament; (ii) a short flexible hook treated as a torsion spring about the motor axis whose hydrodynamics can be   neglected~\cite{Shum2012}; (iii)   a helical flagellar filament of the normal left-handed polymeric form whose hydrodynamics is captured with slender-body theory~\cite{Johnson1980}. Motor and filament parameters are chosen to match those of \emph{E.~coli} bacteria~\cite{Darnton2007} (Table S1 in supplementary information).  Each helical filament has a tapered end such that the helix radius is zero at its attachment point to the motor~\cite{Higdon1979}. Flagellar filaments can  rotate but not translate relative to their attachment point on  the cell body and while the rotation about the helix axis is imposed by the motor, any further rotations relative to the body are solved for. We neglect hydrodynamic interactions between the cell body and flagellar filaments but include steric interactions to prevent   filaments  from entering the body. For each hook, we use $\theta$ to denote the tilt angle between the normal to the cell body at the motor location   and the axis of the flagellar filament (i.e.~when $\theta=0$ the filament is normal to the cell body). The  restoring elastic moment  imposed by each motor on its flagellar filament  is modelled as  torsion spring of spring constant $K = EI/\ell_h$, where $EI$ and $\ell_h$ are the  bending rigidity  and length of the hook respectively~\cite{Son2013}. The magnitude of the   restoring moment is thus given by $K |\theta|$ and the elasticity of the hook acts to align the helix axis with the normal to the cell body.  The computational model solves for the instantaneous   positions  of the flagellar filaments   and for the swimming velocity, $\Ub$, and angular velocity, $\Omegab$, of the cell body   as a function of the hook stiffness.

\begin{figure*}[t!]
\centering 
 \includegraphics[width = 0.91\textwidth]{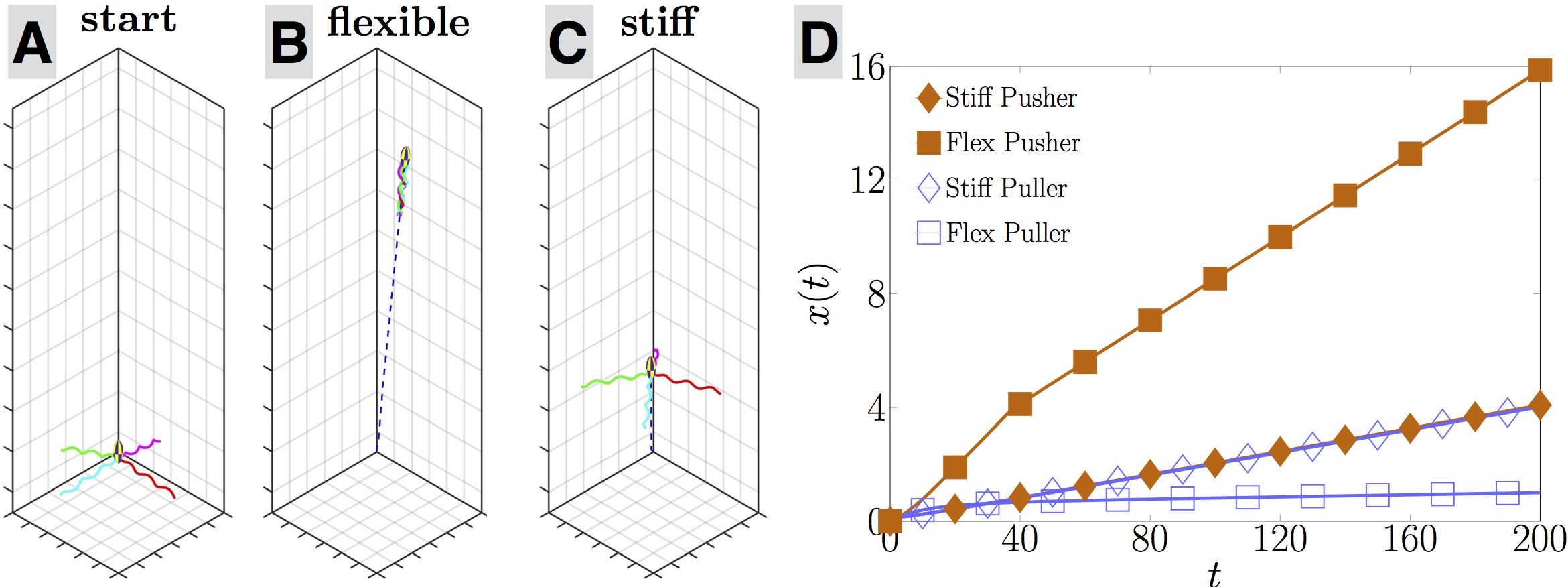}
\caption{
Swimming of a bacterium   with $N_f=4$ flagella
with a flexible vs.~stiff hook.  
A: Initial position and conformation of  each  cell; 
B: Pusher cell with flexible hook at $t=200$ (times scaled by rotation rate of flagella); 
C: Pusher cell  with stiff  hook at $t=200$; 
D: Distance travelled by each swimmer (nondimensionalised  by the pitch of the helical filaments)  as a function of time for four different swimmers: stiff (diamonds)  vs.~flexible hook (squares) and   pusher (filled symbols) vs.~puller (empty).
}
\label{FourFlag}
\end{figure*}

\subsection*{Pusher bacteria with flexible hooks undergo a swimming instability}

Examining the results of our computational model uncovers a remarkable  elastohydrodynamic instability,   illustrated in Fig.~\ref{FourFlag} in the case of  $N_f=4$   flagella, the average number of flagella on an {\it E.~coli} cell \cite{Turner2000}. The motors are  positioned symmetrically around the surface of the cell body. We start the computations with each flagellar filament tilted at some small angle away from the normal to the surface and march the system forward in time while tracking the position of the cell  in the laboratory frame and  of the flagellar filaments relative to the cell body. Associated movies are available in supplementary information.

In Fig.~\ref{FourFlag}A-C we illustrate the trajectory of a pusher bacterium (i.e.~a cell with flagellar filaments undergoing normal CCW rotation) with two different hook stiffnesses  over a time scale $t=200$ (time  nondimensionalised by the rotation rate of the flagella). While both   start at the same location (A),  the cell with the flexible hook ($K=0.1$) ends up with their flagellar filaments all wrapped in the back   and is able to swim five times as fast  (B) as the  stiff-hooked cell ($K=100$) whose flagellar filaments have remained in the same splayed configuration (C). 
This is   quantified in Fig.~\ref{FourFlag}D where we plot the net distance travelled   as a function of time  (scaled by the pitch of the helix). The  cell with a flexible hook  (filled square) swims consistently  faster than the stiff one (filled diamond). If alternatively we reverse the direction of rotation of the flagella to rotate in the clockwise (CW) direction, the cell becomes a puller and does not  transition to fast swimming for neither a flexible hook (empty squares) nor a stiff one (empty diamonds). Note that the two stiff cases (pushers and pullers; diamonds) have identical swimming magnitude, a consequence of  the kinematics reversibility of Stokes flows \cite{Purcell1977}. Importantly, the transition to fast swimming for flexible pusher bacteria does not occur smoothly with changes in the hook stiffness but instead it takes place at a critical dimensionless value of $K_c\approx 1$ (nondimensionalised using the viscosity of the fluid, the pitch of the helical filament and the frequency of rotation).    Above $K_c$, all flagella remain normal to the cell ($\theta\approx 0$)  leading to negligible swimming while below $K_c$,  all flagella wrap behind the cell ($|\theta|\approx \pi/2$) leading to a net locomotion.

\begin{figure*}[t]
\centering
\begin{tabular}{cc}
    \includegraphics[width=0.93\textwidth]{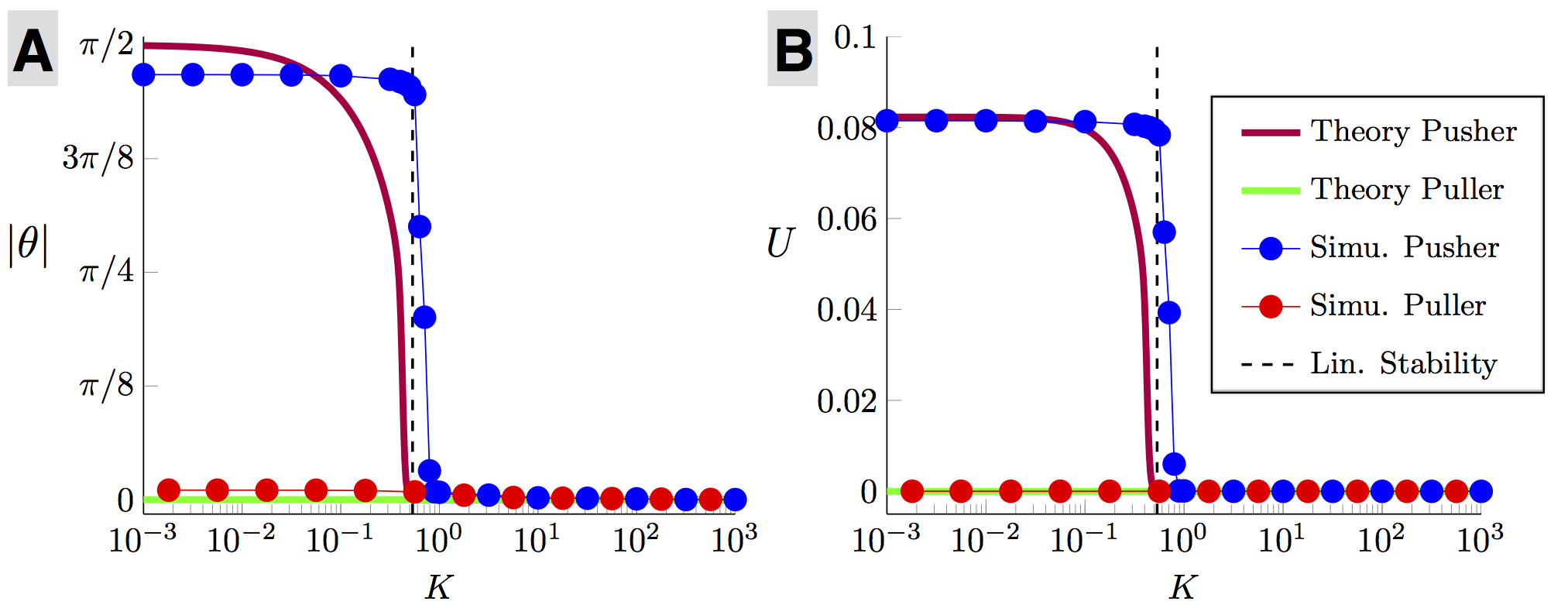}
\end{tabular}
\caption{Steady-state flagella tilt angles ($|\theta|$, A) and lab-frame swimming speeds ($U$, B) for the full computational model of Fig.~\ref{BactDiagram}A
 with two flagella (symbols and thin lines) and for the simple active filament model of Fig.~\ref{BactDiagram}B (thick lines) as a function of the dimensionless hook  spring constant, $K$. Light green line and light red symbols: puller bacterium for which the non-swimming state is always stable; Dark red line and dark blue symbols: pusher bacterium which undergoes a transition to swimming for $K<K_c$. The dashed line shows the critical spring constant predicted theoretically, $K_{c}\approx 0.53$.}
\label{KHelix}
\label{CspRod}
\end{figure*}

This sharp transition does not originate from  a buckling instability of the hook  which   is only modelled here at the level of a torsional spring~\cite{Son2013}. Instead, the instability arises from  the two-way coupling between the conformation of the flagella and cell locomotion.  To unravel the physics of this instability, we consider in more detail the   case of a spherical cell body and two flagella, which is the minimum configuration able to show the instability while capturing the same physics as geometrically-complex cases. The steady-state computational results in this case are shown in the main part of Fig.~\ref{KHelix}   (symbols and thin lines)  for the angle $\theta$   between the axis of the flagellar filaments and the cell body (A) and for the net lab-frame  swimming speed $U$ of the cell (B). While the flagella conformation of  puller bacteria  is   independent of the hook stiffness and leads to zero  swimming (light red circles), pusher cells  clearly display  a sudden jump to a wrapped conformation and a net locomotion for a hook stiffness below   $K_c\approx 0.79$ (dark blue circles).

 \subsection*{Analytical model of the elastohydrodynamic instability}

The observed dynamics can be captured by an analytical   model  demonstrating that swimming occurs as the result of a linear elastohydrodynamic instability. Consider the simple geometrical model illustrated in Fig.~\ref{BactDiagram}.  Two straight active filaments of length $\ell$  are symmetrically  attached  on either side of a spherical cell body of radius $a$  and are tilted at an angle $\pm\theta$ away from the  body surface normal, $\N$.  
 Each filament, elastically attached to the cell body via a hook modelled as a torsion spring of  stiffness $K$,  pushes on the cell along their tangential direction with  propulsive force density $f \uvec{t}$    which results in the  swimming of the bacterium  with velocity $U\uvec{y}$ (see Fig.~\ref{BactDiagram} for all notation).  For CCW motion,  the   propulsion forces point towards the cell body  ($f<0$) and the cell is a pusher. In contrast, for CW motion,   the propulsive forces point away from the cell   ($f>0$) and the swimmer is a puller.

The   swimming speed ($U$) and the rate of change of the conformation of the filaments ($\dot \theta$) may be obtained by enforcing force and moment  balance. Using $c\pa$ and $c\pe$ to denote the drag coefficients for a slender filament moving parallel and perpendicular to its   tangent respectively, the balance of  forces on the whole   cell in the direction of swimming, $\hat {\bf y}$, is written as 
\begin{align}\label{SSrod}
- 6\pi\mu a U -2\ell U\left(c\pa\sin^2\theta  + c\pe\cos^2\theta\right)\\
\nonumber +\dot{\theta}\ell^2c\pe\cos\theta = 2f\ell\sin\theta ,
\end{align}
where the first two terms (the terms on the first line) are due to the drag on the cell body and on  the active filaments  due to swimming, the third term is the drag on the filaments due to rotation and the last term is the total propulsive force acting on the cell.  

The second equation comes from the balance of moment on each active filament,  written in the $\hat {\bf z} = \ \hat {\bf x}\times \hat {\bf y}$ direction at the attachment point on the cell surface as
\begin{equation}\label{eq:moments}
-\frac{\ell^3}{3}c\pe\dot{\theta}+\frac{\ell^2}{2}Uc\pe\cos\theta - K\theta= 0,
\end{equation}
where the first term is the hydrodynamic moment due to rotation of the filament, the second is the hydrodynamic moment due to the swimming drag and the last term is the elastic restoring   moment from the hook    acting to return the filament to its straight  configuration. Combining Eqs.~\eqref{SSrod} and~\eqref{eq:moments} leads to  the   evolution equation for $\theta$
\begin{align}\label{eqdottheta}
\left(\frac{\ell^3}{3}c\pe-\frac{c\pe^2\cos^2\theta\ell^4}{12\pi\mu a +4\ell(c\pa\sin^2\theta +c\pe\cos^2\theta)}\right)\dot{\theta}=\\
\nonumber
\frac{-f\ell^3\sin\theta\cos\theta c\pe}{6\pi\mu a +2\ell(c\pa\sin^2\theta +c\pe\cos^2\theta)} - K\theta.
\end{align}

When the elastic moment dominates, the straight configuration  $\theta=0$ is the only steady state, associated with no swimming. If instead the elastic moment is negligible, the swimming states with $\theta=\pm\pi/2$ become possible equilibria. 

To examine how a variation of the hook stiffness allows  transition from one state to the next, we  solve Eq.~\eqref{eqdottheta} numerically   with the appropriate flagellar filament values for a wild type swimming \emph{E.~coli} cell and using the   magnitude of $f$ leading to agreement with the full computations   at zero hook stiffness.  We start with small perturbations around $\theta=0$ and compute  the long-time steady state of Eq.~\eqref{eqdottheta}, with results illustrated in  Fig.~\ref{CspRod} for both pusher (dark red line) and puller (light green line).  Puller cells  never swim for any value of the hook stiffness, and the straight configuration $\theta=0$ is always stable. In contrast, pushers cannot swim for hooks stiffer than a critical value   but undergo a sudden transition to direct swimming for softer hooks, in excellent agreement with the computations of the full two-flagella case (symbols in Fig.~\ref{CspRod}).

 The sudden transition to swimming for a critical hook stiffness can be predicted analytically by linearising Eq.~\eqref{eqdottheta} near  the equilibrium at $\theta=0$, leading to  
\begin{equation}\label{approxtheta}
\left(\frac{ 4\pi\mu ac\pe\ell^3  +  \frac{1}{3}c\pe^2\ell^4 }{ 12\pi\mu a+ 4c\pe\ell }\right) \dot{\theta}\approx - \left(K + \frac{ fc\pe\ell^3}{6\pi\mu a+2c\pe\ell }\right)\theta.
\end{equation}
If $f$ is positive (puller) then the configuration with $\theta=0$, which is associated with no swimming $U=0$,  is always linearly stable to small  perturbations for any value of $K$. In contrast, pushers with $f < 0$  are  linearly unstable for $K<K_{c} $ such that the right-hand side of Eq.~\eqref{approxtheta} becomes positive, i.e.~$ K_{c} = -{fc\pe\ell^3}/{(2c\pe\ell+6\pi\mu a)}$. A linear elastohydrodynamic instability enables therefore pusher bacteria with sufficiently-flexible hooks to dynamically transition to an asymmetric conformation ($\theta\neq0$) with net swimming ($U\neq 0$). Note that the simple theoretical model (linear stability and numerical solution of Eq.~\ref{eqdottheta})   predicts a critical dimensionless stiffness of $K_c\approx 0.53$,  in    agreement with   the computations for the full bacterium model, $K_c\approx 0.79$.

\section*{Discussion}

How does this swimming instability affect real  bacteria? We first note that for the instability to be  relevant, the rotary motors need to be spatially distributed around the organism and therefore the instability would  not occur if the rotary motors were all located around the  same position on the cell body. Lophotrichous  bacteria whose multiple flagella are   positioned at the pole of the cell (for example, {\it Helicobacter pylori}) would therefore not be subject to this  instability, but peritrichous bacteria such as \emph{E.~coli} and   \emph{Salmonella enterica} would. 

 By examining past measurements on the bending stiffness of peritrichous bacterial hooks, we next discover that  swimming bacteria are safely on the unstable side, explaining their ability to swim despite the presence of spatially distributed motors. The strength of the torsion spring in our model is given by $K = EI/\ell_h$, where $EI$ is the bending rigidity of the hook and $\ell_h$ its  length.  
A recent study measured the hook flexibility for    different  species of peritrichous bacteria by extracting and staining the flagellar hooks and 
using electron microscopy to  observe their deformations due to thermal fluctuations ~\cite{Sen2004}. In this study they found that  \emph{E.~coli} and   {\emph{Salmonella enterica}} had similar  hook bending stiffness in the range  $EI\approx1.6-4.8\times10^{-28}~\rm{Nm}^{2}$ while singly flagellated bacteria can have  stiffer hooks~\cite{Son2013}. Re-dimensionalising our   computational   results above using the viscosity of water (1~mPas) and the pitch (2.22~$\mu$m) and frequency (110~Hz) of {\it E.~coli} flagella \cite{Darnton2007}, we obtain 
$K_c\approx 9.6\times 10^{-19}$~Nm. Using  a hook length  $\ell_h\approx55$~nm~\cite{berg03},  
we therefore predict a critical hook stiffness of $EI \approx 5.5\times10^{-26}~\rm{Nm}^{2}$.   The hook stiffness  of peritrichous bacteria is thus   two orders of magnitude smaller  than the critical value for the instability.
   
In summary, we showed theoretically and computationally that pusher peritrichous bacteria can swim by exploiting an elastohydrodynamic   instability while pullers never can. This instability is due to the bending rigidity of the hooks and is different from the buckling instability displayed by polar bacteria~\cite{Son2013}.  The physics of this instability lies in the feedback between the conformation of the flagella and the swimming of the cell. Flagellar filaments create propulsive forces which propel the cell forward. In the frame of the moving cell, the filaments experience hydrodynamic moments aligning them with the direction of swimming. The rigidity of the hook balances these hydrodynamic moments and below a critical rigidity, an elastohydrodynamic instability transitions the filaments from a splayed state to a conformation where they are gathered behind the cell. Our results rationalise the ability of real peritrichous bacteria to swim by showing that they are designed to undergo a successful transition to swimming after each tumble.

\section*{Methods}
We give here a brief outline of the computational model, with  all details found in supplementary information. 
 Our computational model solves for the instantaneous   positions  of the flagellar filaments   and for the swimming velocity, $\Ub$, and angular velocity, $\Omegab$, of the cell body  by enforcing mechanical equilibrium at all time  (inertia is irrelevant at the scale of bacteria).   At low Reynolds numbers, the balance of hydrodynamic    forces and moments  for the whole cell at its centre    leads to a linear relationship between the swimming kinematics of the cell and the angular velocities of each filament,  denoted  by $\omegaf_i$ for $i^{\rm th}$ filament, of the form
\begin{equation}
 \begin{pmatrix}
  \Ub\\
  \Omegab
 \end{pmatrix}
=
\sN {\boldsymbol \Upsilon}_i\cdot \omegaf_i,
\end{equation}
where the tensors ${\boldsymbol \Upsilon}_i$ depend on the hydrodynamic resistance of each individual component of the cell and on their geometrical arrangements.  

The rotation rate of each filament has a prescribed value   along its  helical axis and we need two additional equations to solve for the other two components. This is obtained by examining the local balance of moments.  In the frame of a filament, the swimming velocity and rotations are experienced as background flows which act to tilt the flagellum away from the normal to the motor while the hook applies an elastic restoring moment. The hydrodynamic  moment acting on 
 filament $i$ may be written as $
{\bf L}_{i}= {\boldsymbol \Gamma}_i\cdot\Ub +  {\boldsymbol \Lambda}_i\cdot\Omegab +  {\boldsymbol \Delta}_i\cdot\omegaf_i,$ 
 where the tensors ${\boldsymbol \Gamma}_i$, ${\boldsymbol \Lambda}_i$ and ${\boldsymbol \Delta}_i$ dependent on the geometry and relative configuration of the flagellar filament and cell body, and are proportional to the fluid viscosity.   
If we use the unit vector $\uvec{N}_i$ to denote  the direction normal to the cell body surface at the location of the motor and if  $\uvec{k}_i$ is the unit vector along     the axis of the filament 
(see  Fig.~\ref{BactDiagram}, top)  then the restoring elastic moment acts along the unit vector  $\hat{\bf H}_i=\hat{\bf k}_i\times\hat{\bf N}_i$ and the balance between the hydrodynamic moment and restoring elastic moment from the hook is written for all times as
 \begin{equation}
{\bf L}_{i}\cdot\uvec{H}+K\theta_i = 0.
\end{equation}
Finally, we assume that there is no elastic resistance for the filament   to move in the  direction $\hat{\bf J}_i=\hat {\bf k}_i\times \hat{\bf H}_i$ perpendicular to both ${\bf k}_i$ and $\hat{\bf H}_i$  and thus the final moment equation is  ${\bf L}_{i}\cdot\uvec{J}=0$.

\subsection*{Acknowledgements}
We thank Lyndon Koens for useful discussions. This project has  received funding from the European Research Council (ERC) under the European Union's Horizon 2020 research and innovation programme  (grant agreement 682754 to EL) and from   the EPSRC (EER).

\end{document}